# Competing magnetic interactions in the orthorhombic GdNiAl$_3$

S. Nallamuthu[1], K. Arun[1], Sergej Ilkovic[2], Ivan Curlik[2], Marian Reiffers[2], R. Nagalakshmi[1*]

[1]Department of Physics, National Institute of Technology, Tiruchirappalli 620 015, India
[2]Faculty of Humanities and Natural Sciences, Presov University, Presov, Slovakia

**ABSTRACT**

Magnetization and heat capacity measurements of ternary rare earth intermetallic compound GdNiAl$_3$ demonstrate para to ferromagnetic transition at Tc=165.5K. In addition multiple short range magnetic transitions observed below Tc are suggestive of competing interactions in this compound. As a result of this a weak Griffiths phase type behaviour is observed in the paramagnetic region. This complex behaviour is rather supported by the random orientation of Ni centered tricapped trigonal prisms with additional Al atoms in the structure. Heat capacity and resistivity data display an interesting peak at 72 K, which is highly unaffected by magnetic fields up to 90KOe.

Keywords: Magnetization; ferromagnetic; Heat capacity; magnetic entropy.

*Corresponding author: Tel.: 04312503615; e-mail: nagaphys@yahoo.com, nagalakshmi@nitt.edu (R.Nagalakshmi),

# 1. INTRODUCTION

Ternary intermetallic gadolinium based systems have shown a wealth of interesting issues such as spin glass, magnetostriction, giant magnetoresistance, magnetocaloric effect and high magnetic refrigeration capacity [1-3]. Gd has two interesting specific features associated with the high spin and zero angular momentum. Among the rare earths, $Gd^{3+}$ has $^8S_{7/2}$ ground state that leads to high effective coupling around room temperature and large magnetic moment, which in turn largest magnetic entropy. The L = 0 for 4f orbital implies no magneto-crystalline anisotropy resulting from crystal electric effects in Gd [4,5]. In the ternary metallic Al-Gd-Ni system fourteen intermetallic compounds with different crystallographic structures were reported [6] and hence it is a well explored system. The nearest neighborhoods of Gd and Ni ions, as well as the distances Ni-Ni, Gd-Gd, Gd-Ni are different in each compound. This leads to different states of Ni ions and also influences the Gd-Ni interaction in these compounds [7]. In many rare-earth based Ni compounds, nickel atoms do not carry magnetic moment because of charge transfer of Gd conduction electrons to the 3d band [8, 9]. It is expected in such a system that we can observe at low temperatures, very different magnetic structures, such as ferromagnetism, ferrimagnetism or antiferromagnetism and competing magnetic interactions as well. Also multiple magnetic phases are common in Gd –Ni-Al systems such as $GdNiAl_4$ [10], $GdNiAl_2$ [11], $GdNi_3Al_{16}$ [12] are of special interest as they lead to exotic complex magnetic behaviours.

$GdNiAl_3$ belongs to the family of ternary intermetallic compound R-Ni-Al systems of $RTX_3$ type. The different structures of aluminides are located on the line between the binary RNi and Al [13]. From the structure report of $GdNiAl_3$, it is understood that the compound adopts the orthorhombic $YNiAl_3$ type structure [14]. Hitherto, no detailed investigations on the magnetic behaviours of this particular stoichiometry are available. This prompted us to undertake the systematic studies on the structural, magnetization, heat capacity and resistivity of stoichiometric polycrystalline $GdNiAl_3$ sample for the first time. GdNiAl3 compound reveals multiple magnetic transitions exhibiting complex magnetic behaviour. These successive transitions are reminiscent of competing interaction in this system which may stem from the distortion/randomness in the structure.

## 2. EXPERIMENTAL METHODS

The specimen of nominal composition GdNiAl$_3$ were synthesized from the elements of high purity of Gd (99.9%), Ni (99.99%) and Al (99.999%) in the stoichiometric 1:1:3 ratio by arc-melting on a water-cooled copper crucible with a tungsten electrode under purified argon atmosphere. The resulting alloy button was turned over and remelted several times to ensure homogeneity. The ingots were annealed at 750ºC under vacuum in quartz ampoule for 10 days. To confirm the phase purity and stoichiometry of the annealed sample powder X-ray diffraction using Cu-Kα radiation and scanning electron microscope (SEM)-energy dispersive X-ray spectroscopy analyses were carried out.

DC magnetization was measured as a function of temperature M (T) and magnetic field M (H) using Quantum Design physical property measurement system (PPMS) upto 90 kOe. For zero field cooled (ZFC), the sample is cooled below T$_C$ in the absence of a magnetic field upto 50K. Then small magnetic field is applied to increase the temperature. For field cooled (FC) the temperature of the sample is lowered in field below T$_C$. The ac susceptibility measurements were done as a function of frequency from 100 to 1000 Hz at probing fields of 2.5Oe using the Quantum Design physical property measurement system (PPMS) from 1.8K to 300 K. Heat capacity was also measured using PPMS relaxation technique down to 1.8K with external magnetic field upto 9T. Four-probe resistivity measurements were carried out in longitudinal geometry using a homemade set up along with Oxford superconducting magnet system from 1.5 to 300 K in zero and magnetic fields up to 80 kOe.

## 3. RESULTS AND DISCUSSION

### 3.1. X-ray diffraction

Room-temperature powder x-ray diffraction (XRD) pattern was recorded using a Bruker AXS diffractometer (Cu Kα) radiation, 2θ ranges from 10° to 80° with a step size of 0.02 and 5 s/step counting time. The collected powder pattern was used for phase identification of the given compound using the FULLPROF software package [15]. Rietveld refinement of the data along with XRD pattern of GdNiAl$_3$ is shown in Fig. 1. Lattice parameters obtained by Rietveld refinement a = 8.1654Å, b=4.0692Å and c = 10.6652Å are in good agreement with the previously reported data [13]. The analysis indicates that the annealed samples are in single phase with the YNiAl$_3$-type orthorhombic structure (space group Pnma) [14]. There is an impurity GdAl$_2$ phase which shows less than wt. 4%, and we believe that it does not make any substantial contribution to major properties. It is intriguing in the reported structure [13], that the excess Al atoms (site Al3) are situated inside the deformed cubic coordinated Ni centred polyhedra consisting of R and Al atoms. Further each Ni centred GdAl trigonal prisms shown in Fig. 2b. are located in random direction. This may be essential for the contribution of uncertainty in the structure which in turn be the source of complex magnetic behaviour leading to competition of magnetic interactions in this novel compound. The micrograph of SEM image with EDAX spectrum of GdNiAl3, confirms the 1:1:3 stoichiometry. The crystallite sizes are generally between 1 – 10 µm, which shows the good quality of the sample.

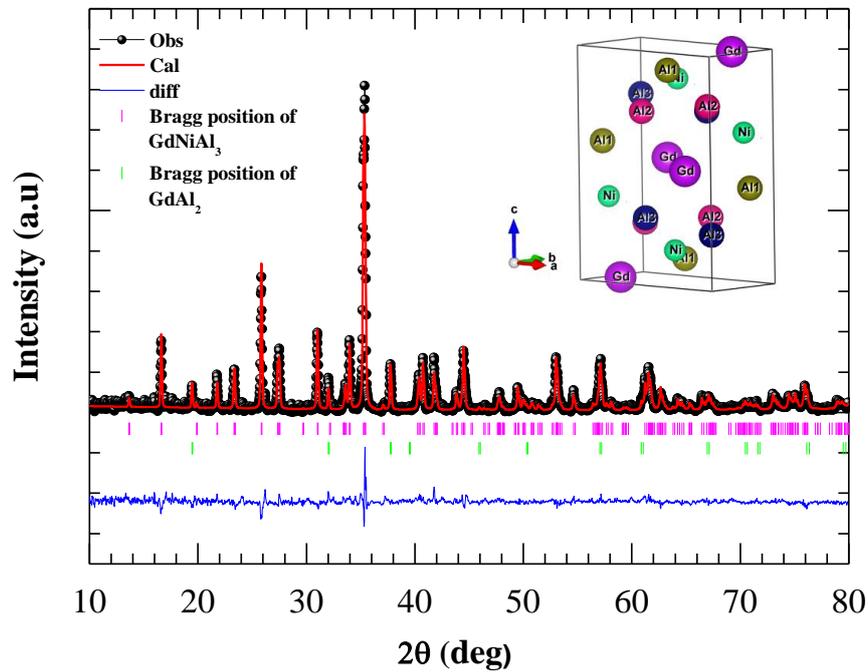

**Fig. 1.** Powder X-ray diffraction pattern of GdNiAl$_3$ along with Rietveld refinement using GSAS program.

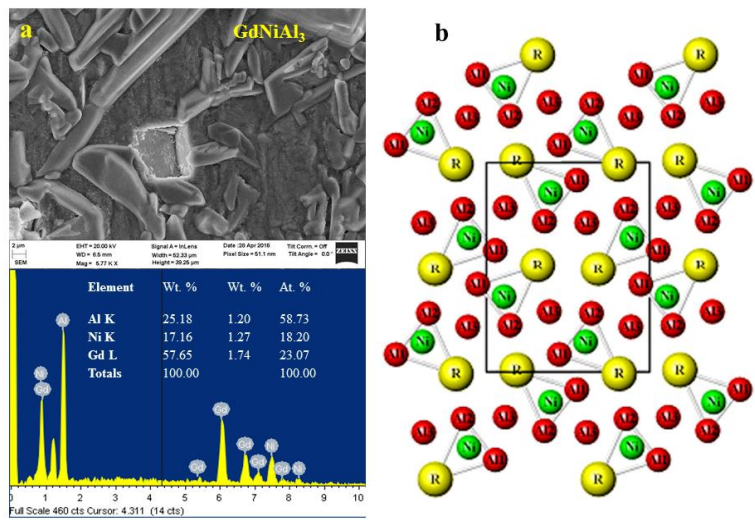

**Fig. 2.** (a). The SEM image along with EDAX spectrum. (b). Ni centred GdAl trigonal prisms in RNiAl$_3$ (R = Gd) [Ref 13].

## 3.2. Magnetic susceptibility and magnetization studies

The temperature dependence of dc magnetic susceptibility ($\chi$ = M/H) and inverse susceptibility ($\chi^{-1}$) of GdNiAl3 measured in different applied fields such as 1 kOe and 5 kOe are shown in Fig. 3a, 3b and 3c. The susceptibility exhibits a slope change at $T_C$ =165.5K and tends to saturate on further decreasing the temperature and thereby revealing the presence of paramagnetic to ferromagnetic transition at the Curie temperature $T_C$ = 165.5K, which is defined as the inflection point in the ($d\chi^{-1}/dT$) curve. For an external applied field of 5 kOe the position of slope change remains unaffected whereas the susceptibility displays subtle humps representing several short range ordering below $T_C$ at T = 65 K and 11 K (Fig. 3b). By increasing the magnetic field of 30kOe (inset Fig. 3b shown from 50-300K), the slope change at $T_C$ gets smoothened and the kink at 65K becoming prominent. Thus it may be related to the change in magnetic structure by the application of magnetic field. These features indicate the possibility of competing interactions, thereby exhibiting the metastable ground state. Similar type of complex magnetic behaviour has been observed in $R_2PdSi_3$ [16], $Ho_2Mn_3Si_5$ [17], $R_2Mn_3Si_5$ [18]. The inverse susceptibility ($\chi^{-1}$) of GdNiAl3 at 1kOe exhibits a slope change around T = 165.5 K and tends to saturate on further decreasing the temperature below $T_C$, which is the characteristic of a ferromagnetic transition. The inverse molar susceptibility is consistent with Curie-Weiss like behaviour in the temperature range 240-400K and fitting this temperature range with Curie Weiss law

$$\chi = \frac{C}{(T-\theta_p)} \quad (1)$$

Where θp is the Curie temperature and C is the Curie constant which can be expressed in terms of effective moment as

$$C = \frac{\mu_{eff}^2 x}{8} \qquad (2)$$

Where $x$ is the number of rare-earth atoms per formula unit. The solid lines in Fig. 3b are fits to Equation.1. The value of effective magnetic moment ($\mu_{eff}$ = 7.97$\mu_B$) obtained is equal to the expected theoretical value for $Gd^{3+}$ free ions (7.94 $\mu_B$) and positive $\theta_p$ (35 K) values confirm that the dominant interactions are of ferromagnetic in nature.

However, the inverse susceptibility does not follow Curie-Weiss law above $T_C$ and below 240 K; rather a convex behaviour is seen. Hence the deviation from Curie-Weiss law above $T_C$ to 240K may due to the existence of some short range ferromagnetic correlations above $T_C$. This deviation is suppressed for a high field of about 30 kOe (inset of Fig. 3c), as this short range correlations are masked by paramagnetic signal. Thus the whole system does not develop a long range magnetic ordering. This feature may strongly hint us the presence of Griffith's Phase [19-22]. The competition of magnetic interactions also favours the observation of Griffith's phase.

To probe further the magnetic behaviour of the sample below $T_C$, chosen sets of magnetic isotherms were measured from 3 – 50 K in the increasing field upto 50 kOe and shown in Fig. 4b. A slight immediate increase in magnetization and presence of a weak hysteresis support the argument that the primary interactions are ferromagnetic. However, a non-saturating behaviour and smaller than expected magnetic Gd moments in $GdNiAl_3$ suggest that the magnetic state is not fully ferromagnetic. This is an indicative of the anisotropic behaviour in some Gd based compounds [23, 24]. At very low temperature and upto 50K, isothermal thermal magnetization curves indicate a weak field induced spin flop like metamagnetic transition around 30kOe which is a representative of antiferromagnetic correlations of short range. The critical magnetic field for magnetic transition is determined from the maximum of dM/dH curve. No spin flop type transition is observed in M (H) curve for 110K, 150K and 180 K.

Fig. 4 (c) shows the Arrott plot displaying the nonlinear variation of $M^2$ as function of H/M without intercepting the $M^2$ axis. This suggests the absence of spontaneous magnetization. Examination of Arrot plot shows that nonlinear isothermal curves centered at 30 kOe may be associated with spin flop type metamagnetic transition, which is symbolic of antiferromagnetic ordering [25-27]. Positive gradient values in Arrot plots near critical temperature have been ascribed to second order phase transition.

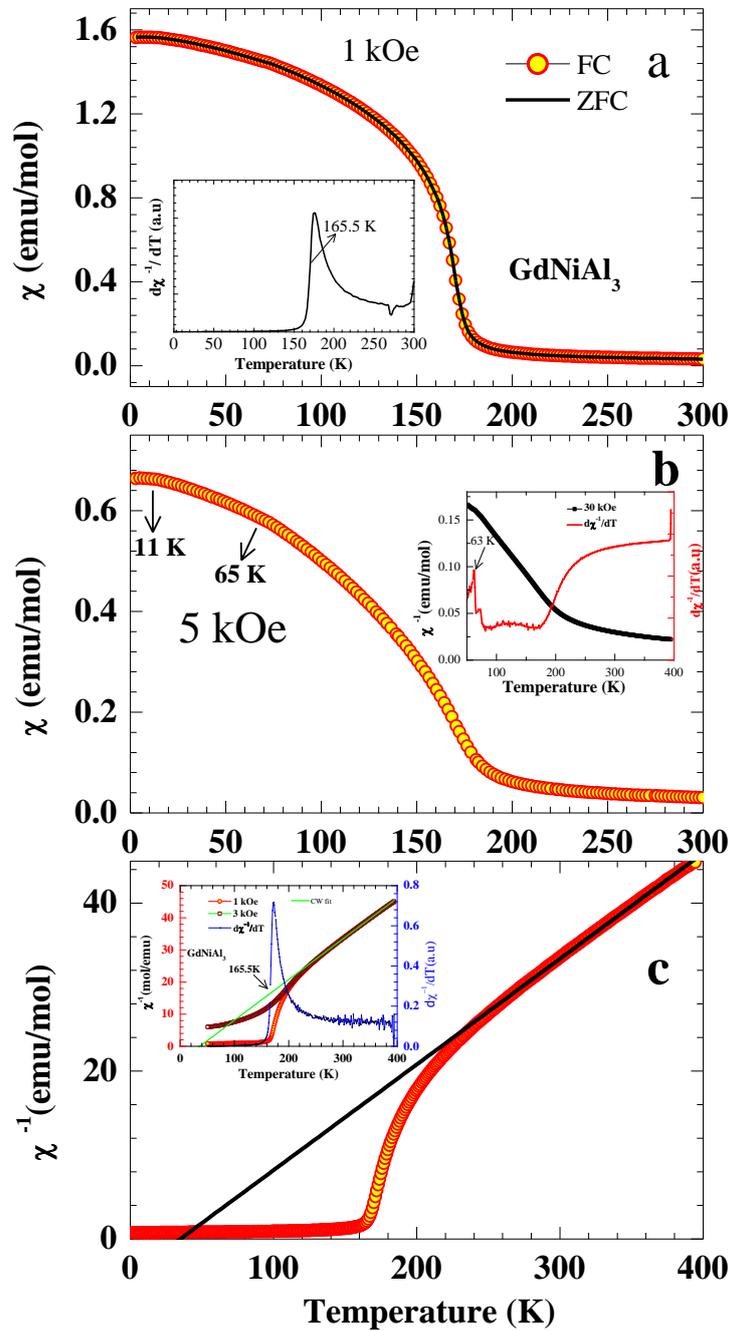

**Fig. 3.** (a). and (b). The temperature dependence of zero field cooled (ZFC) and field cooled (FC) dc magnetic susceptibility of GdNiAl$_3$ in the temperature range 3–300 K measured in different magnetic fields. (c). The inverse of the magnetic susceptibility ($\chi^{-1}$) of GdNiAl$_3$ in the temperature range 3 – 300 K measured in H = 1 kOe. The inset shows that derivative of inverse susceptibility (d$\chi^{-1}$/dT). The solid lines represent the fits to Curie-Weiss relation.

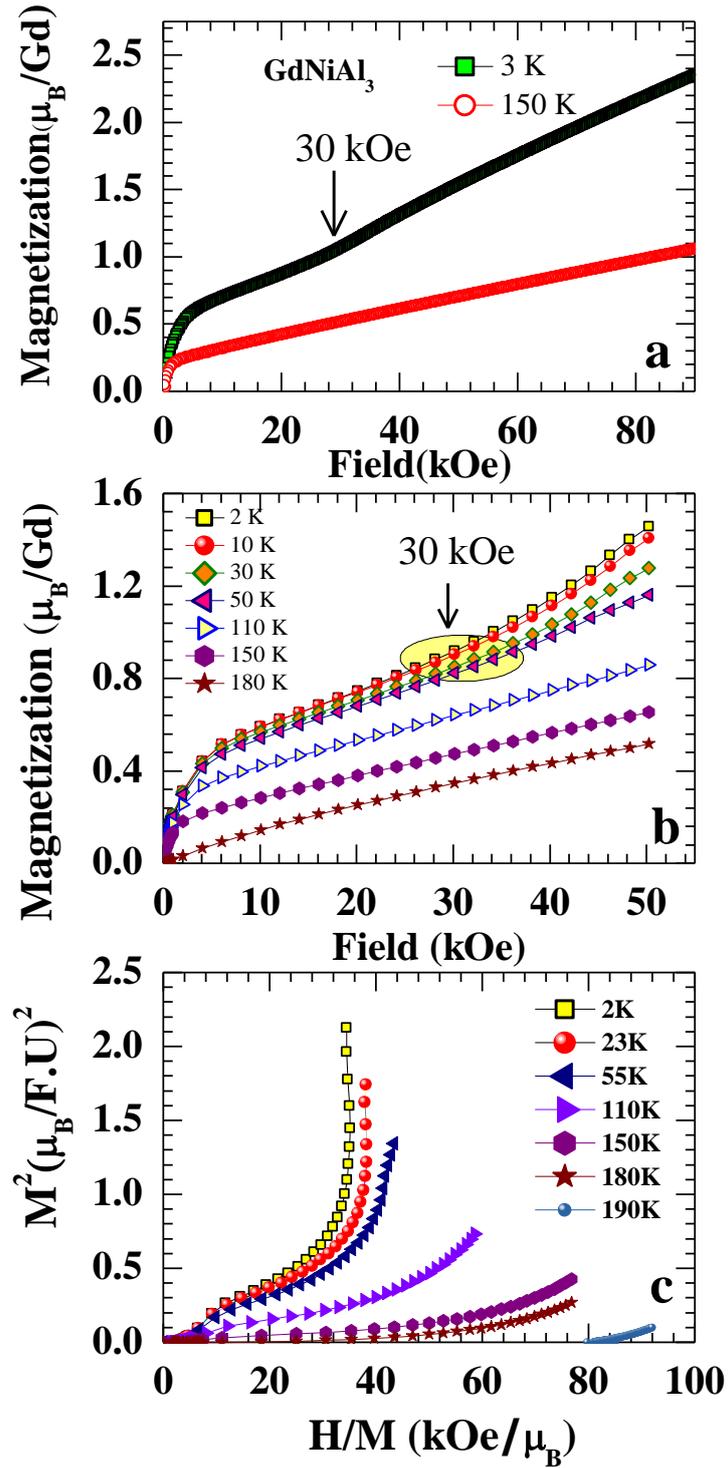

**Fig. 4.** (a). Isothermal magnetization curves of GdNiAl$_3$ as a function of magnetic field upto 90 kOe at 3 K and 150 K. (b). Isothermal magnetization of GdNiAl$_3$ at selected temperatures upto 50 kOe. (c). Arrott plot of magnetic isotherms for GdNiAl$_3$ (H ≤ 50 kOe) at various temperatures

### 3.3. AC Magnetic Susceptibility

The nature of magnetic interactions was further studied by performing AC susceptibility measurements at different frequencies and in order to elicit more information on phase transition of this compound, the temperature dependence of magnetization was carried out in ZFC and FC modes at low field H = 50 Oe and is shown in Fig. 5a. The pair of FC- ZFC curves measured in 50Oe bifurcates at Tc= 165.5K marking the beginning of irreversible phenomena which is manifested by deviation of FC from ZFC curve of $\chi_{dc}$. This is typical for a system with competing magnetic interactions. Such thermomagnetic irreversibility often arises in ferromagnets when the anisotropy energy is equivalent to the interaction energy leading to domains with narrow walls which indicates the presence of domain wall pinning effect in this compound [28]. Also it is realized in materials having competing interactions, ferromagnetic materials having large anisotropy and in antiferromagnets having large random orientation of crystallites or randomly quenched spin ordering such as spin-glass state [29, 30]. So as to clarify the bifurcation in the terms of spin glass, we performed AC susceptibility measurements for GdNiAl$_3$ compound. Fig. 5b and 5c shows the temperature variation of the in phase (real part $\chi'$) and out of phase (imaginary part $\chi''$) components at various frequencies ($\nu = \omega/2\pi$) of 100, 504 and 997 for an AC driving field of 5 Oe and DC field H$_{DC}$ = 0Oe.

$\chi_{ac}$ can be written as

$$\chi = \frac{dM}{dH} = \chi' + i\chi'' \qquad (3)$$

Where $\chi'$ is the real part related to the reversible magnetization process, and $\chi''$ is the imaginary part related to irreversible magnetization process and energy absorbed from field. The slope change at T$_C$ = 165.5K is frequency independent in $\chi'_{ac}$ versus T curve which rules out the possibility of spin glass behaviour in this compound. On the other hand the out of phase $\chi''_{ac}$ does not show any variation in the position of T$_C$ and displaying a moderate frequency dependence below T$_C$. Small kinks at 69 K and 3.5 K are perceptible for frequencies 100 Hz and 504 Hz. These observations shows that slope change at Tc = 165.5K in real and imaginary parts suggests the paramagnetic to ferromagnetic transition while the finite value of kinks appeared below T$_C$ in $\chi''$ may be associated with emerging successive antiferromagnetic correlations. Thus the competing antiferromagnetic and ferromagnetic ground states lead to complex behaviour in this compound.

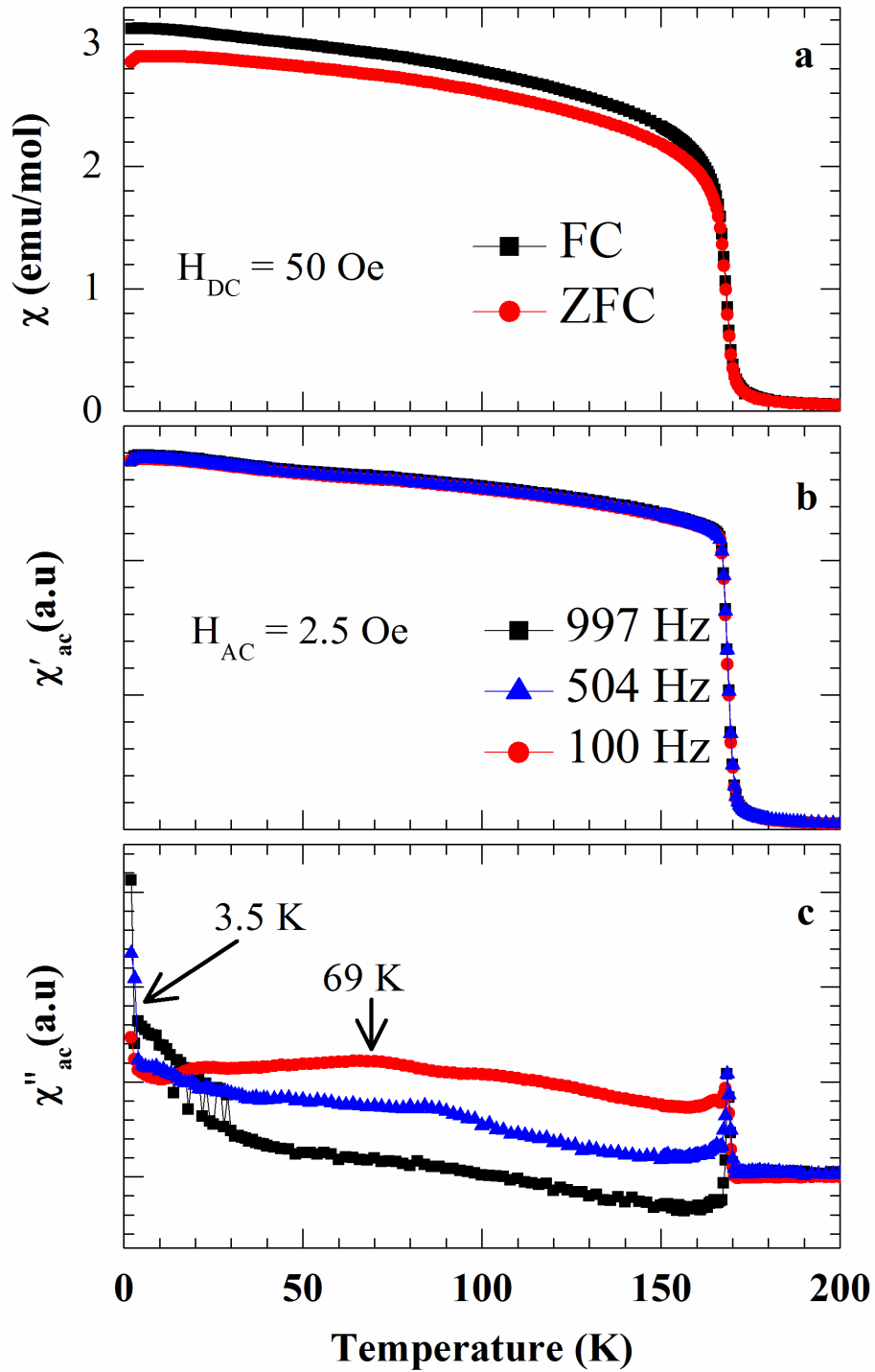

**Fig. 5.** DC and AC magnetic susceptibility of GdNiAl$_3$. DC magnetic susceptibility measured under ZFC and FC conditions for H$_{DC}$ = 0 Oe. AC susceptibility measured at different frequencies such as 100, 504 and 997 Hz for H$_{AC}$ = 2.5 Oe.

## 3.4. Specific heat

The temperature dependent specific heat of GdNiAl$_3$ measured from 2 K to 300 K at various fixed fields such as 0Oe, 100Oe. 10kOe, 20kOe and 30kOe, 50 kOe and 90 kOe is shown in Fig. 6a and 6b. The high temperature peak observed in the C$_P$ (T) curves is a general characteristic of intermetallic compounds showing the onset of long range magnetic ordering. In the case of GdNiAl$_3$, the heat capacity data plotted in Fig. 6a show a similar occurrence of multiple magnetic realignment in the ordered state. The sharp peak at 165.5 K in the heat capacity correlates with the peak exhibited by susceptibility at around that temperature. This T$_C$ is in agreement with the ferromagnetic ordering as seen in magnetization data.

The broad peak observed at T$_C$ =165.5K is gradually suppressed and shifts to higher temperature with the application of magnetic field is evocative of ferromagnetic behaviour and is of second order magnetic phase transition. Furthermore, some multiple anomalies are observed below T$_C$ in heat capacity data at 72.5 K, 65 K, 11 K and 3.5 K respectively (Fig. 6b). The application of magnetic field shifts the peak at T$_C$ to high temperatures which is the usual case of ferromagnet. However the remaining anomalies at 65 K, 11 K and 3.5 K are moved to low temperatures with respect to magnetic field. This behaviour provides evidence for antiferromagnetic correlations in this compound. A striking feature of this compound is that the anomaly at 72.5 K is completely impervious to magnetic field and also not appeared in thermomagnetic curve (compare with the Fig. 6b and Fig. 6a). One would presume from heat capacity and resistivity data, the sharp transition at 72.5 K could be associated with some structural changes in GdNiAl$_3$. This has to be further studied and confirmed from temperature dependent X-ray diffraction measurements. Hence, the AFM magnetic state is more likely below structural transition temperature followed by a paramagnetic to ferromagnetic transition at T = 165.5 K. Thus the heat capacity data clearly reveals the complex behaviour of this compound. Thus, competing interactions might have been originated from the randomness in the structure, as outlined in Fig. 2b.

The magnetic contribution (C$_{4f}$) of GdNiAl$_3$ is estimated by subtracting the specific heat of nonmagnetic YNiAl$_3$ (C$_{4f}$ = C (GdNiAl$_3$) - C (YNiAl$_3$)). The C$_{4f}$ data shows a broad maximum below T$_{ord}$, which could be associated with magnetic contribution. The temperature dependence of magnetic entropy (S$_{4f}$) calculated from C$_{4f}$ of GdNiAl$_3$ has been deduced by using the relation $S_{4f}(T) = \int \frac{C_{4f}}{T} dT$ , as shown in upper inset of Fig. 6b. The estimated magnetic entropy (S$_{4f}$) is reaching maximum of 23.2 J/mol K$^2$ at ordering temperature T$_C$ = 165.5 K. This value is higher than theoretical value of S$_{4f}$ = Rln (2J+1) =

17.28 J/mol K² expected for Gd³⁺ (J = 7/2) ion. The complete entropy has been released at the ordering temperature $T_C$ indicates the spins are fully randomized. At high temperature (T = 300 K), $C_P$ = 119 J/mol K (inset of Fig.6), which is close to the expected classical Dulong-Petit higher value of $C_V$ = 3NR =124.5 J/mol K (N is number of atoms per formula unit and R is the molar gas constant).

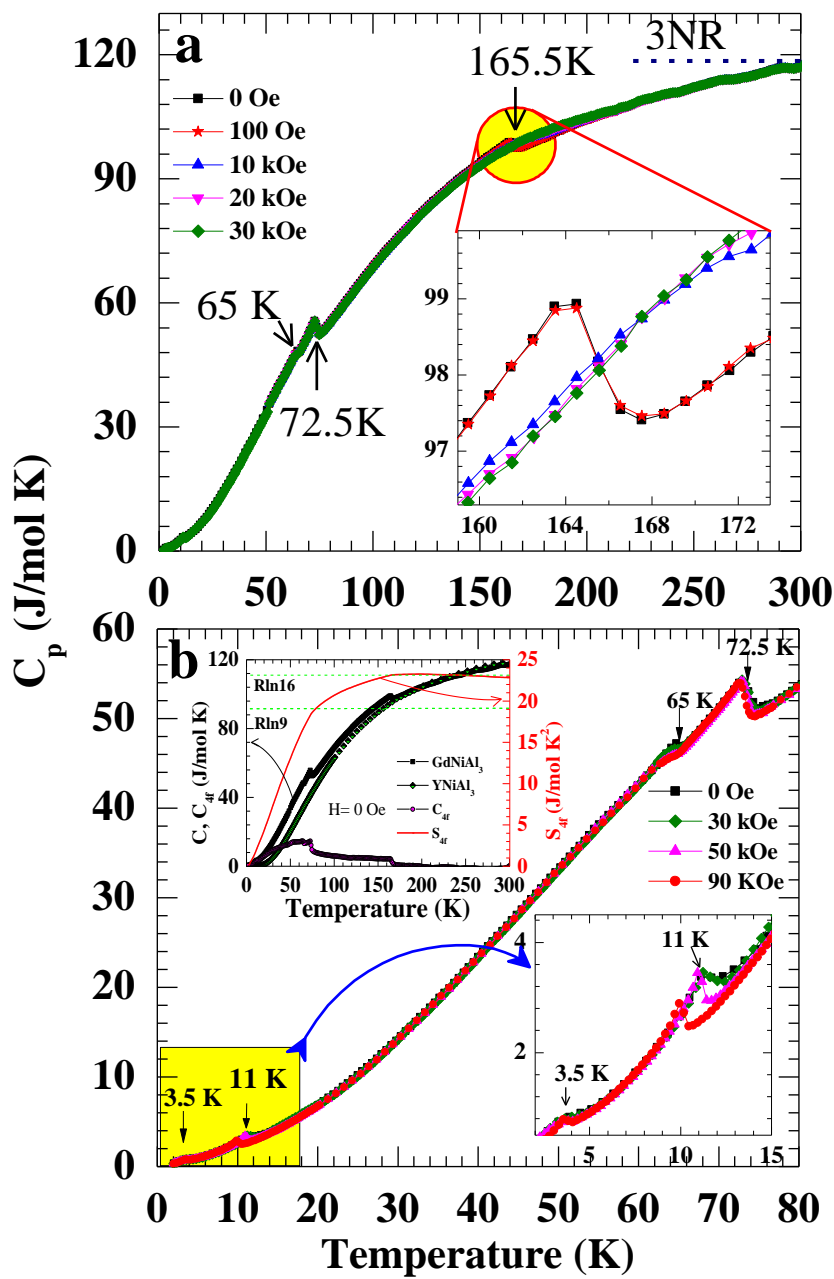

Fig. 6. Temperature dependent specific heat of GdNiAl₃ measured in the temperature range (a). 2–300 K upto 30 kOe. (b) 2–80 K upto 90 kOe. The lower inset shows the expanded view. The upper inset of (b) shows magnetic heat capacity ($C_{4f}$) and entropy ($S_{4f}$).

## 3.6 Electrical resistivity

Fig. 7 shows the temperature dependences of electrical resistivity for GdNiAl$_3$ at zero and in field of 80kOe. The compound exhibits normal metallic behaviour at high temperatures. There is a resistivity drop around T=72.5 K, which coincides with the heat capacity data. Also it acts as a precursor for structural phase transition. At low temperatures, the Fermi liquid (FL) behaviour $\rho = \rho_0 + AT^n$ ($\rho_0$ is residual resistivity, n = 2) is found to be valid. We have fitted to the low temperature region less than 10 K to FL fit which describe the values of $\rho_0$ = 68 µΩ cm. The residual resistivity ratio (RRR) [$\rho$(300 K)/$\rho_0$] determined as 6.10, suggesting the good quality of the sample. Application of magnetic field of 80KOe invokes no considerable changes in resistivity data.

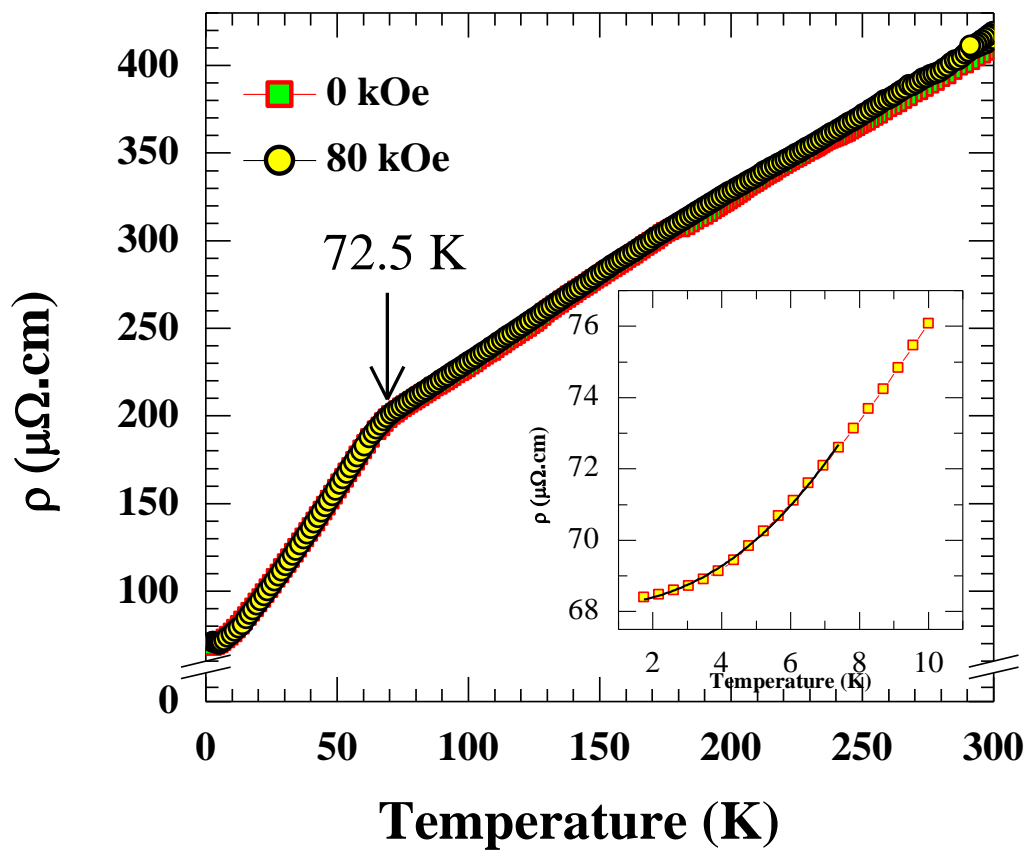

Fig. 7. Electrical resistivity $\rho$ (T) of GdNiAl$_3$ measured in zero magnetic field and 80 kOe. Insets shows fit to the $\rho = \rho_0 + AT^2$.

## 4. CONCLUSION

We have investigated the magnetic, thermodynamic and transport properties of polycrystalline GdNiAl$_3$, which is isotypic to YNiAl$_3$ using suite of experimental techniques for the first time. GdNiAl$_3$ found to exhibit ferromagnetic behaviour at T$_C$ =165.5K. The divergence in ZFC and FC along with the ac susceptibility confirm the domain reorientation behaviour in this compound. Multiple magnetic field sensitive transitions below Tc manifest the existence of competing ferro and antiferromagnetic interactions as confirmed from magnetic and heat capacity data. We presume that the randomly distributed spins in the unsystematic arrangement of rare earth trigonal prisms having Ni at the centre with additional Al atoms might have induced the complex alignment of magnetic moments and hence competing interactions in GdNiAl$_3$. The peak at 72K is completely insensitive to magnetic fields as consistent from heat capacity and resistivity drives us to ascribe it for structural transition. Further experimental investigations of the compound by temperature dependent XRD is expected to yield an insight for this anomalous transition. In summary, our results suggest a complex nature of the multiple transitions in the novel GdNiAl$_3$ compound.

## ACKNOWLEDGEMENTS

The authors (R.N) thank Department of Atomic Energy (DAE), Board of Research in Nuclear Sciences (BRNS) Govt. of India for supporting this work under DAE Young Scientists Research Award (No: 2010/20/37P/BRNS/2513). The author (S.N) also thank BRNS for awarding JRF in the project.

# REFERENCES

1. K A Gschneidner Jr, V K Pecharsky and A O Tsokol, Rep. Prog. Phys. 68 (2005) 1479–1539.

2. Sachin Gupta, K.G. Suresh and A.K. Nigam, Journal of Alloys and Compounds. 586 (2014) 600–604.

3. M. Napoletano, F. Canepa, P. Manfrinettia and F. Merlo, J. Mater. Chem., 10 (2000) 1663-1665.

4. M. Angst, A. Kreyssig, Y. Janssen, J.-W. Kim, L. Tan, D. Wermeille, Y. Mozharivskyij, A. Kracher, A. I. Goldman, and P. C. Canfield, Phys. Rev. B. 72 (2005) 174407-13.

5. C. Song, W. Good, D. Wermeille, A. I. Goldman, S. L. Bud'ko, and P. C. Canfield, Phys. Rev. B. 65 (2005) 172415-4.

6. P. Villars and L. D. Calvert, Pearson's Handbook of Crystallographic Data for Intermetallic Phases, Vol. 1 (American Society of Metals, Cleveland, OH, 1994), p. 858.

7. M. Coldea, V. Pop, M. Neumann, O. Isnard, L.G. Pascut, Magnetic properties of Al–Gd–Ni orthorhombic compounds, Journal of Alloys and Compounds, Volume 390, 2005, 16-20.

8. H. R. Kirchmager, E. Burzo and Landolt-Borstein New Series III/19d2 (1990) 248.

9. V. Pop, M. Coldea, M. Neumann, S. Chiuzbaian, D. Todoran, X-ray photoelectron spectroscopy and magnetism of Gd3Ni8Al, Journal of Alloys and Compounds, 333, 2002, 1-3..

10. G.A. Stewart, W.D. Hutchison, A.V.J. Edge, K. Rupprecht, G. Wortmann, K. Nishimura Y. Isikawa. Journal of Magnetism and Magnetic Materials 292 (2005) 72–78

11. Renwen Li, Z. Ma, E. Agurgo Balfour, H. Fu, Y. Luo, Critical behavior study in GdNiAl2 intermetallic compound, Journal of Alloys and Compounds, 658, 2016, 672-677.

12. M. Coldea, V. Pop, M. Neumann, S.G. Chiuzbaian, D. Todoran, Journal of Magnetism and Magnetic Materials, 242–245, 2002, 864-866.

13. Svitlana Pukas, Oksana Matselko, Roman Gladyshevskii, Chem. Met. Alloys. 3 (2010) 35-41

14. R. E. Gladyshevskii, E. Parthe, Acta Cryst. C. 48 (1992) 229-232.

15. Rodríguez-Carvajal, J. Recent Developments of the Program FULLPROF, in Commission on Powder Diffraction (IUCr). Newsletter (2001), 26, 12-19

16. A. Szytuła, M. Hofmann, B. Penc, M. Ślaski, Subham Majumdar, E.V. Sampathkumaran, A. Zygmunt, Journal of Magnetism and Magnetic Materials, 202, 1999, 365-375

17. R. Nirmala, A. V. Morozkin, Jagat Lamsal, Z. Chu, V. Sankaranarayanan, K. Sethupathi, Y. Yamamoto, H. Hori, W. B. Yelon, and S. K. Malik. Journal of Applied Physics 105, 07A721 (2009)

18. R. Nirmala , V. Sankaranarayanan , K. Sethupathi , A.V. Morozkin. Journal of Alloys and Compounds 325 (2001) 37–41

19. F. Issaouia,  M.Bejar, E.Dhahri, M.Bekri, P.Lachkar, E.K.Hlil, Physica B. 414 (2013) 42–49.

20. Wanjun Jiang, X. Z. Zhou, and Gwyn Williams, Y. Mukovskii and K. Glazyrin, Phys. Rev. B. 76 (2007) 092404-4.

21. J. Deisenhofer, D. Braak, H.-A. Krug von Nidda, J. Hemberger, R. M. Eremina, V. A. Ivanshin, A. M. Balbashov, G. Jug,6 A. Loidl, T. Kimura, and Y. Tokura, Phys. Rev. Lett. 95 (2005) 257202-4.

22. Krishanu Ghosh, Chandan Mazumdar, R. Ranganathan and S. Mukherjee, Scientific Reports 5, (2015), 15801

23. A. Garnier, D. Gignoux, N. Iwata, D. Schmitt, T. Shigeoka, F.Y. Zhang, Anisotropic metamagnetism in $GdRu_2Si_2$, Journal of Magnetism and Magnetic Materials, 140–144, 1995, 899-900

24. Andrea Marcinkova, Clarina de la Cruz, Joshua Yip, Liang L. Zhao, Jiakui K. Wang, E. Svanidze, E. Morosan, Journal of Magnetism and Magnetic Materials, 384, 2015, 192-203.

25. P. M. Shand, C. C. Stark, D. Williams, M. A. Morales, T. M. Pekarek, D. L. Leslie-Pelecky, J. Appl. Phys. 97 (2005) 10J505-3.

26. A. Aharony and E. Pytte, Phys. Rev. Lett. 45 (1980) 1583-1586.

27. S. Jia, N. Ni, G. D. Samolyuk, A. Safa-Sefat, K. Dennis, Hyunjin Ko, G. J. Miller, S. L.  Bud'ko, P. C. Canfield, Phys. Rev. B. 77 (2008) 104408-14.

28. G Liang, F Yen, S Keith and M Croft, J. Magn. Magn. Mater. 314 (2007) 52–59.

29. K. Binder, and A. P. Young, Rev. Mod. Phys. 58 (1986) 801-976.

30. P A Joy, Anil Kumar and S K Date, J. Phys. Condens. Matter. 10 (1998) 11049–11054.